\documentclass[fleqn,twoside]{article}
\usepackage{espcrc2}
\usepackage{psfig}

\usepackage{graphicx}
\usepackage[figuresright]{rotating}

\title{Probing the Dark Ages with the Square Kilometer Array}

\author{C.L. Carilli\address{National Radio Astronomy Observatory,
    Socorro, NM, 87801, USA, ccarilli@nrao.edu},
S. Furlanetto\address{Caltech, Pasadena, CA, USA},
F. Briggs\address{Australia Telescope National Facility,
Sydney, Australia},
M. Jarvis, S. Rawlings\address{Astrophysics, Dept. of Physics, 
Keble Road, Oxford OX1 3RH, UK},
H. Falcke \address{ASTRON, P.O. Box 2, 7990 AA Dwingeloo, The Netherlands}
}
       
\begin{document}

\begin{abstract}

The epoch of reionization (EoR) sets a fundamental benchmark in cosmic
structure formation, corresponding to the formation of the first
luminous objects that act to ionize the neutral intergalactic medium
(IGM). Recent observations at near-IR and radio wavelengths imply that
we are finally probing into this key epoch of galaxy formation at $z
\ge 6$.  The Square Kilometer Array (SKA) will provide critical
insight into the EoR in a number of ways. First, the ability of the
SKA to image the neutral IGM in 21cm emission is a truly unique probe
of the process of reionization, and is recognized as the next
necessary and fundamental step in our study of the evolution of large
scale structure and cosmic reionization. Second, study of HI 21cm
absorption toward the first radio loud objects probes small to
intermediate scale structure in the neutral 'cosmic web', as well as
HI in the first collapsed structures (proto-disks and mini-halos). And
third, the incomparable sensitivity of the SKA allows for the study of
the molecular gas, dust, and star formation activity in the first
galaxies, as well as the radio continuum emission from the first
accreting massive black holes. Such objects will be obscured at
optical wavelengths due to absorption by the neutral IGM.

\end{abstract}

\maketitle

\section{Introduction}

After recombining at $z \sim 1000$, the intergalactic medium (IGM)
remains neutral until the formation of the first luminous sources in
the universe.  These act to reionize the universe, leading to the
fully ionized IGM seen today (neutral fraction $\sim 10^{-5}$). The
epoch of cosmic reionization (EoR) therefore sets a fundamental
benchmark in cosmic structure formation, corresponding to the
formation of the first luminous objects (star forming galaxies and/or
accreting massive black holes). 

The last few years has seen a revolution in our understanding of the
EoR, in particular, the first observational constraints on the
redshift for the EoR. The discovery of the Gunn--Peterson absorption
trough in the spectra of the most distant quasars ($z>6$),
corresponding to Ly $\alpha$ absorption by the neutral IGM, implies
that we have finally probed into the EoR \cite{White03}.  These data
show that the neutral fraction of the IGM decreases rapidly over a
short cosmic time, from $>10^{-3}$ at $z=6.3$, to $<10^{-4}$ at $z =
5.8$. A rapid transition is supported by a number of lines of
evidence, including: (i) studies of the thermal state of the IGM at
high redshift via high $z$ Ly$\alpha$ forest observations
\cite{Hui03}, (ii) observations of cosmic Stromgren spheres and
surfaces around the highest redshift QSOs
\cite{Walter03,Wyithe03,Mesinger04}, and (iii) $\Lambda$CDM galaxy
formation models normalized to known high $z$ galaxy populations
\cite{Harford03}.  However, recent CMB large scale polarization
measurements by WMAP suggest a significantly ionized IGM at $z \sim
10$ to 20 \cite{Kogut03}. Taken together, current observational
constraints suggest that cosmic reionization had a fairly complex
history, extending from $z=6$ to 20 \cite{Cen02}.

While the discovery of the GP effect, and the large scale polarization
of the CMB, are clearly monumental achievements in observational
cosmology, both techniques have limitations.  For the GP effect, the
IGM becomes opaque for neutral fractions $\ge 10^{-3}$, such that the
technique loses it's diagnostic power once the IGM becomes a small
fraction neutral. For the CMB, the large scale polarized signal is an
integral measure of the universal optical depth to Thompson
scattering, and can be fit with myriad reionization scenarios
\cite{Gnedin04a}. The SKA will play a pivotal role in determining the
epoch, and physics, of cosmic reionization in two ways: (1) through
direct observations of the neutral IGM via the HI 21cm line, and (2)
through continuum and line observations of the first luminous objects
-- stars, galaxies and black holes.

The ability of the SKA to image the neutral IGM in 21cm emission is a
truly unique probe of the process of reionization, and is recognized
as the next necessary and fundamental step in our study of
cosmic reionization. Likewise, study of HI 21cm absorption toward the
first radio loud objects probes small to
intermediate scale structure in the neutral 'cosmic web', as
well as HI in the first collapsed structures (proto-disks and
mini-halos). Observations of the HI 21cm line are a direct
probe of the evolution of the neutral IGM, and provide key diagnostics
on the ionization and thermal state of the IGM, the formation of large
and small scale structure, and the nature of the first luminous
objects (AGN or stars).

The incomparable sensitivity of the SKA allows for the study of the
molecular gas, dust, and star formation activity in the first
galaxies, as well as the radio continuum emission from the first
accreting massive black holes.  It is important to keep in mind that,
due to the on-set of GP absorption at $z \sim 6$, observations of
objects within the EoR are limited to wavelengths longer than 0.9
micron. Hence study of the first galaxies and AGN is the exclusive 
realm of  near-IR to radio astronomy (and X-rays). 

The study of the evolution of the neutral IGM, and the first luminous
objects, has been selected as one of the key science programs for the
SKA.  Details of these studies can be found in contributions
by Furlanetto,  Carilli, Blain, Falcke, and Jarvis in this volume.
We review this material herein.

\section{Observing the HI 21cm line from the Epoch of Reionization}

\subsection{Tomography}

The study of the redshifted HI 21cm line at meter wavelengths with
the SKA provides a direct probe of the pristine
neutral IGM within the EoR.  Unlike studies of the Ly$\alpha$ forest,
21 cm tomography does not suffer from saturation when the neutral
fraction is large, so it is ideal for studying the middle and early
stages of reionization.  Moreover, because it is a spectral line
observation, it provides a complete time history of reionization
rather than an integral constraint, such as the optical depth to
electron scattering of the CMB.  The major difficulty with the
technique - the weak signals - makes this an ideal project for the
SKA. 

The first consideration of detecting the 21cm signal from the EoR
entailed searching for a global step in the background temperature due
to the transition from a neutral to ionized IGM \cite{Shaver99}.
Given this is a full-sky measurement, the sensitivity is independent
of the telescope area. The most recent treatment of this problem can
be found in \cite{Gnedin04b}.  This is a difficult observation,
requiring detecting a change in the overall background spectral index
at the 0.1$\%$ level, and the difficulty is excerbated if the EoR
extends over a large redshift range.

A more interesting application is to map out variations in the neutral
hydrogen density, either across the sky or in frequency space.  Such
``tomography'' studies allows one to reconstruct the time evolution of
reionization in an entirely unique manner.  The idea is to measure the
fluctuating brightness temperature of the IGM gas, which depends in
turn on the local overdensity, neutral fraction, and spin temperature.
In general we expect three distinct observational epochs.  Before the
first luminous sources form, we should see weak fluctuations from the
``proto-cosmic web'' and virialized gas in dark matter ``minihalos''
\cite{iliev02,iliev03}.  After these sources turn on, Ly$\alpha$
photons flood the universe and render the entire IGM visible against
the CMB.  Depending on the thermal history of the gas, this era can be
seen in either absorption or emission, so this era can be used to
study the first luminous sources in the universe.  Once the IGM is
heated to well above the CMB temperature (either through radiation or
shock heating), temperature fluctuations no longer affect the
brightness temperature and we see only fluctuations in the density and
ionized fraction.

\begin{figure*}[htb]
\includegraphics*[width=6in]{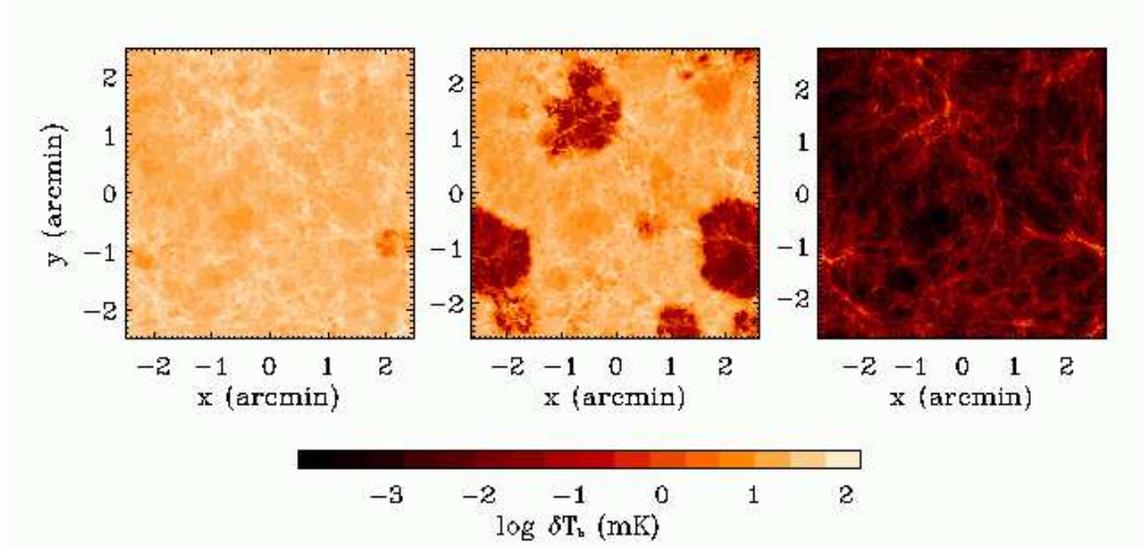}
\vspace{-0.5in}
\caption{ The brightness temperature of the 21 cm transition at
several redshifts, as predicted by the ``late reionization''
simulation analyzed in \cite{Furlanetto04} (see contribution
by Furlanetto, this volume).  Each
panel corresponds to the same slice of the simulation box (with width
$10 h^{-1}$ comoving Mpc and depth $\Delta \nu=0.1$ MHz), at
$z=12.1$, 9.2, and 7.6, from left to right. The three epochs shown
correspond to the early, middle, and late stages of reionization in
this simulation.}
\label{fig:sim}
\end{figure*}

We expect most of the interesting phases of reionization to take place
during this last era.  Str{\" o}mgren spheres surrounding ionizing
sources rapidly dominate the fluctuations (see Fig. \ref{fig:sim}).
However, our current understanding of this process is poor and
measurements of reionization are crucial for advancing our
understanding of this landmark event.  In particular, we would like to
know how the HII regions grow around sources, and which sources are
most responsible for reionization.  The contribution by
Furlanetto (this volume) describes several
theoretical studies of the signals expected during reionization.  In
all models, we expect fluctuations of order 1--10 mK on scales of one
to several arcminutes, with the characteristic scale increasing as
reionization proceeds \cite{Furlanetto04}.  The growth of these
regions, as well as the mean signal, directly measures the entire
ionization history, the morphology of the HII regions, and the path by
which the IGM is ionized (for example, whether voids or dense regions
are ionized first).  The expected signal is thus rich in potential
physical diagnostics of both the IGM and the first generations of
luminous sources.
 
The SKA will be the first instrument with adequate sensitivity to
\emph{image} fluctuations over the entire set of angular scales
relevant to these observations, especially near the beginning of
reionization when the HII regions are small (near or somewhat below an
arcminute in size).  While instruments such as
PAST\footnote{astrophysics.phys.cmu.edu/\~jbp} and
LOFAR\footnote{www.lofar.org} should be able to measure some of the
statistical properties of neutral gas and possibly map the sky on
coarse angular scales, the SKA offers a unique opportunity to image
the evolution of large-scale structures during the EoR (Figure 1).
The expected frequency range of 100 MHz to 200 MHz ($z = 6$ to 13) is
sufficient to place powerful constraints on the ionization history,
though measurements of ``first light'' may require lower frequencies.
One of the major challenges in recovering this signal is foreground
contamination from Galactic synchrotron radiation and extragalactic
point sources.  However, there are several promising techniques to
eliminate this contamination.  Zaldariagga et al. \cite{Zaldarriaga04}
showed that the smooth spectra of these foregrounds allow efficient
cleaning through differencing maps closely spaced in frequency.
Morales \& Hewitt \cite{Morales04} showed that symmetries in the
spectral domain also allow foreground removal.  While technically
challenging, we expect that 21 cm tomography will revolutionize our
understanding of the high-redshift universe.

\subsection{Absorption}

Observing HI 21cm emission from the EoR implies studying large
scale structure (cluster scales and larger).  A number of groups
have recently considered the possibility of studying smaller scale
structure in the neutral IGM by looking for HI 21cm absorption toward
the first radio-loud objects (AGN, star forming galaxies, GRBs).

Carilli et al. \cite{Carilli03} use numerical simulations to predict
the HI 21cm absorption profile of the 'cosmic web' prior to
reionization. For example, for a source at $z = 10$, they predict an
average optical depth due to 21cm absorption of about 1$\%$,
corresponding to the 'radio Gunn-Peterson effect'. They also find
about five narrow (few km/s) absorption lines per MHz with optical
depths of a few to 10$\%$ (Figure 2). These latter lines are
equivalent to the Ly $\alpha$ forest seen after reionization, and
correspond to over-densities evolving in the linear regime ($\delta
\le 10$).  While significant questions remain about simulating the
thermal state of the IGM to such detail, the simple point remains
that, while the Ly$\alpha$ lines from such structures are highly
saturated, the (much) lower 21cm oscillator strength makes the IGM
translucent prior to reionization.

\begin{figure}[htb]
\includegraphics[width=3in]{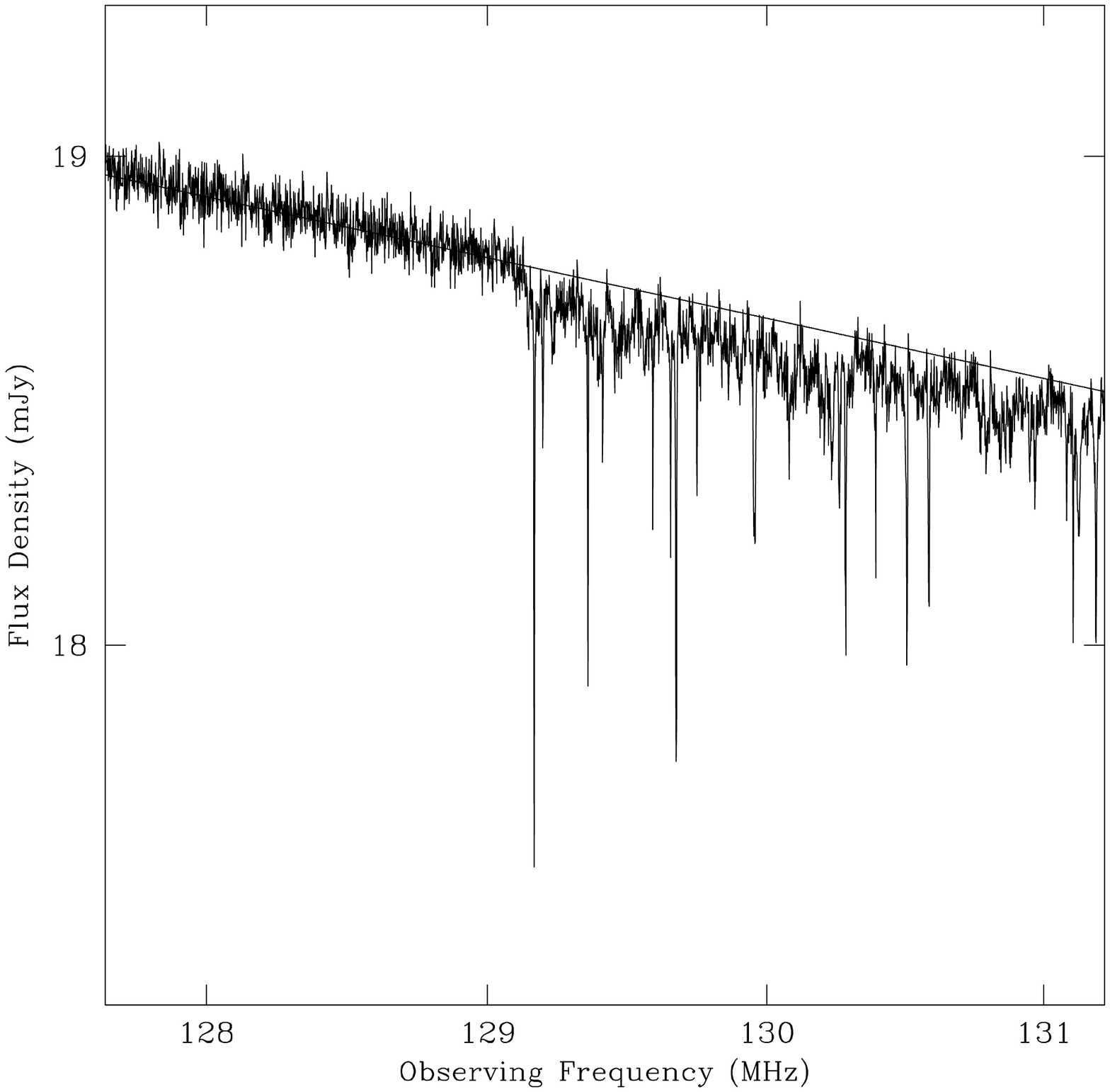} 
\caption{The
simulated spectrum of a source with S$_{120}$ = 20 mJy at $z = 10$
using a spectral model based on the powerful radio galaxy 
Cygnus A  and assuming HI 21cm absorption by
the IGM (see contribution by Carilli, this volume,
\cite{Carilli03}). Thermal noise has been added using the
specifications of the SKA and assuming 10 days integration with 1 kHz
wide spectral channels.  The onset of absorption by the neutral IGM is
seen at 129 MHz, corresponding to the HI 21cm line at $z=10$.}
\end{figure}

It is possible that, due to the lower cosmic Jean's mass prior to
reionization, the majority of the HI is contained in mini-halos,
ie. collapsed structures with $\delta > 100$ and masses $\le 10^7$
M$_\odot$. Furlanetto \& Loeb \cite{Furlanetto03} have considered this
possibility and predict a similar HI 21cm absorption line density due
to gas in mini-halos as that expected for the 21cm forest.

The predicted absorption signal depends on the excitation state of the
HI. This in turn depends on IGM pre-heating, including possible early
reionization (and subsequent recombination), resonant scattering of
Ly$\alpha$ photons, and heating by soft X-rays.  Cen \cite{Cen02}
shows that the presence or absence of HI absorption by mini-halos, or
the cosmic web, would be a telling diagnostic of early IGM heating
mechanisms.

Furlanetto \& Loeb \cite{Furlanetto03} also consider the expected 21cm
absorption profiles for proto-disk galaxies. While such absorption
lines will be rare ($10^4$ times less frequent than the 21cm forest
lines), the optical depths may be large enough that the lines could be
observed toward faint radio sources, in particular, gamma ray burst
radio after-glows within the host galaxy.

An important caveat in these calculations is the assumption of radio
loud sources during the EoR.  This question has been considered in
detail by Carilli et al. \cite{Carilli03} and Haiman et al.
\cite{Haiman04}. They show that current models of radio-loud AGN
evolution predict between 0.05 and 1 radio sources per square degree
at $z > 6$ with $\rm S_{150MHz} \ge 6$ mJy, adequate for EoR HI 21cm
absorption studies with the SKA (see also section 4).

\section{Study of the first luminous objects}

Gunn-Peterson absorption during the EoR ($z \ge 6$) precludes
observations of objects at wavelengths longer than 0.9 micron. Hence
study of the first galaxies and AGN is the exclusive realm
of near-IR to radio astronomy. The last few years has seen
a revolution in the number of objects discovered at $z > 6$ using
near-IR and radio imaging and spectroscopy, including star forming
galaxies \cite{Stanway04,Hu02,Kodaira03}  and AGN \cite{Fan03}. 

The recent discovery of molecular line emission, thermal emission from
warm dust, and radio syncrotron emission, from the most distant QSO
($z=6.4$; \cite{Bertoldi03,Walter03,Carilli04a}), implies very early
enrichment of heavy elements and dust in galaxies, presumably via star
formation, within 0.8 Gyr of the big bang (Figure 3).  Unfortunately,
such experiments stretch current instrumentation to the extreme limit,
such that only rare and pathologic objects are detectable,
ie. hyperluminous IR galaxies with $L_{FIR} > 10^{13}$ L$_\odot$.
The two orders of magnitude increase in sensitivity afforded by the
SKA will enable study of the first 'normal' galaxies
within the EoR in a number of ways. 

The SKA operating at short cm wavelengths ($\sim 20$ GHz) allows for
detailed molecular line imaging of galaxies in the high-$z$
universe. The high redshifts involved mean that the standard mm
molecular transitions will be available for study with the SKA,
eg. observations at 20 GHz correspond to CO 1-0 at $z=5$, CO 2-1 at $z
= 10$, and HCN 1-0 at $z = 3.5$. Such studies reveal the physics and
chemistry of molecular gas reservoirs required for star formation, and
provide a unique probe of gas dynamics and dynamical masses of the
first galaxies \cite{carilli04b}. 
Also, radio continuum studies with nanoJy sensitivity
in the frequency range 1 to 10 GHz will present a dust-unbiased view
of star formation in these systems.

\begin{figure}[htb]
\includegraphics[width=3in]{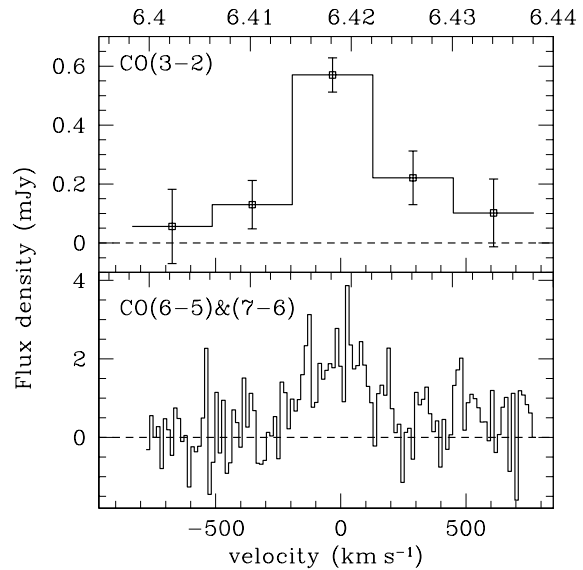}
\caption{The CO line emission from the most distant QSO, 1148+5251 at
$z=6.42$ \cite{Walter03,Bertoldi03b}. The 3-2 line was
observed with the Very Large Array at 47 GHz, while the higher order
transitions were observed with the Plateau de Bure interferometer. The
implied molecular gas mass is $2.2\times 10^{10}$ M$_\odot$.}
\end{figure}

Note that ALMA also has the potential for studying molecular emission
from high redshift objects, but only in systems where very high-order
transitions are excited (eg. observations at 90 GHz correspond to CO
7-6 at $z=8$), and with less sensitivity than SKA, even for thermally
excited lines.  In general, SKA and ALMA will provide complementary
information on the low and high order molecular emission from galaxies
at intermediate redshifts ($z = 3$ to 6), while at the highest
redshifts, ie. into the EoR ($z > 6$), the SKA becomes the premier
instrument for studying the molecular content of early galaxies (see
contribution by Blain, this volume).  For example, Carilli \& Blain
\cite{Blain03} show that in optimal surveys at 20 GHz the SKA will
detect of order 15 sources per hour in CO 1-0 emission at $z \sim 4$.

As a concrete example of the types of objects that might be studied,
Pello et al. \cite{Pello04} have recently presented evidence in the
near-IR for a possible strongly lensed $z \sim 10$ galaxy.  Although
the detection appears to be spurious \cite{Weatherley04}, the proposed
source properties may be considered characteristic of the first
generation of galaxies that will be studied in deep near-IR surveys.
The apparent UV luminosity is $4\times10^{11}$ L$_\odot$.  Making the
standard dust correction for typical high redshift star forming
galaxies (ie. Ly-break galaxies) then implies an apparent (FIR)
luminosity of $2\times 10^{12}$ L$_\odot$. Assuming a line width of
300 km/s, the predicted CO 2-1 velocity integrated flux density is 4.5
mJy km s$^{-1}$ at 21 GHz. This CO emission line could be detected (at
5$\sigma$) with the SKA in 10 minutes. The (non-thermal) radio
continuum emission is about 1 $\mu$Jy at 1.4 GHz, which could be
detected with the SKA in about 15 minutes. And to emphasize
complementarity with future instrumentation, the predicted thermal
emission from warm dust at 250 GHz is 0.6 mJy, which can be detected
with ALMA in a few minutes as well. If we remove the effect of strong
lensing the required integration times increase by about two orders of
magnitude, but that still means only about 20 hours integration to
detect the first galaxies in the universe.  Indeed, Hopkins
\cite{hopkins00} predicts that, in an 8 hr integration with the SKA at
1.4 GHz, one would detect of order 7000 star forming galaxies at $z >
5$ in the 1 deg$^{2}$ FoV of the SKA (see van der Hulst et al this
volume, although cf. Jackson, this volume).

Figure 4 shows the spectrum of the active star forming galaxy Arp 220
at a number of redshifts, compared to the sensitivity of future
instruments.  The SKA is well matched to future instruments such as
ALMA and the JWST in terms of studying the first luminous objects in
the universe, allowing for a panchromatic view of the earliest stages
of galaxy formation. Such a broad view will reveal the complete suite
of physical processes in these systems, including stars, star
formation, the interstellar medium (dust and gas), and the galaxy
dynamics (dark matter), in objects forming within 0.5 Gyr of the big
bang.

Lastly, an important aspect of the molecular line observations of
galaxies within the EoR is that they give the most accurate redshifts
(by far) for the host galaxies. Typical high ionization broad 
metal emission lines from QSOs are notoriously uncertain in terms of
the host galaxy redshifts, with offsets typically on the order of
10$^3$ km s$^{-1}$ \cite{Richards02}, while Ly $\alpha$ emission lines
are affected severely by absorption.  Accurate host galaxy redshifts
are crucial in the calculation of the size of cosmic Stromgren spheres
around objects within the EoR, since these sizes are derived from the
redshift difference between the host galaxy and the on-set of GP
absorption \cite{Walter03}. The sizes of these ionized regions have
been used to constrain the IGM neutral fraction \cite{Wyithe03},
setting a lower limit to the neutral fraction of 0.1 at $z \sim 6.4$,
two orders of magnitude more stringent than the lower limit set by the
GP effect.

\begin{figure*}
\includegraphics*[width=5in]{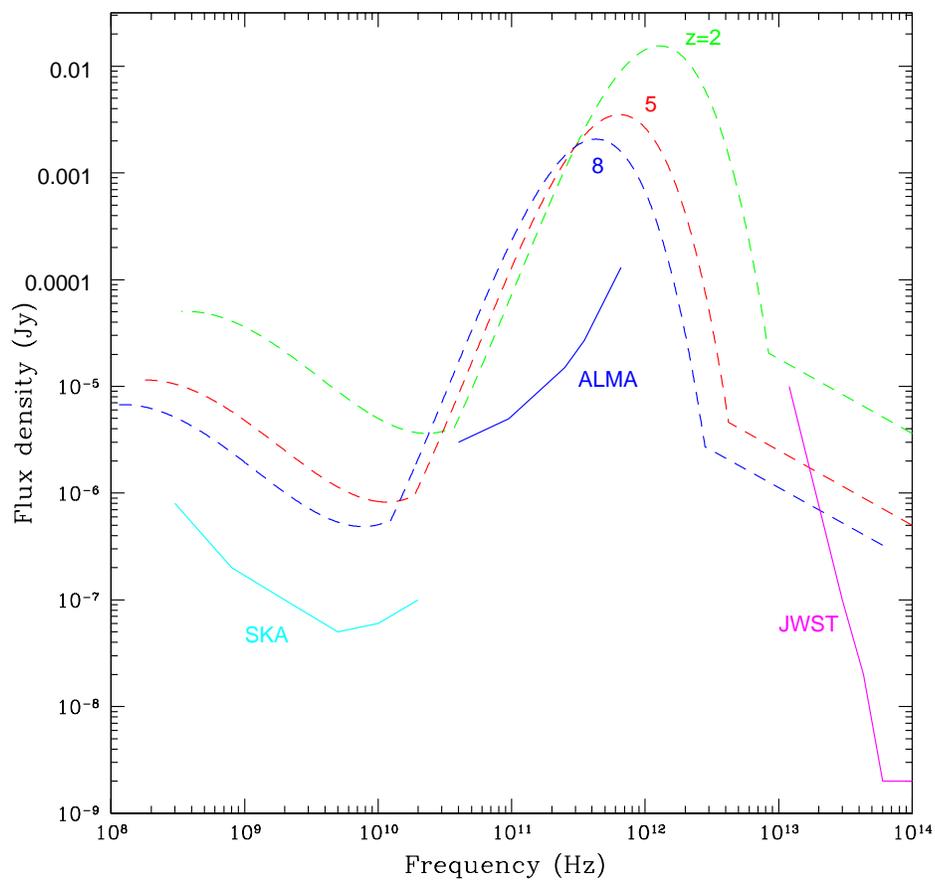}
\caption{The dash lines show the spectrum of the active star forming
galaxy Arp 220 ($L_{FIR} = 1.3\times 10^{12}$ L$_\odot$) at
three  redshifts ($z=2$, 5, 8).
The solid lines show the rms sensitivity of future instruments (in 1hr) at
cm through near-IR wavelengths.}
\end{figure*}

\section{The First AGN}

\subsection{Extended radio sources}

Within the EoR, hierarchical clustering will begin to yield the first
dark-matter halos massive enough to host supermassive black holes
(SMBH). During the major growth phases of these objects, they will
shine as QSOs, although much of the radiation may be absorbed locally
by dense neutral material in the dark-matter halo, or in the more
diffuse neutral IGM. The mass-doubling timescale for an
Eddington-rate-limited SMBH is $\sim 4 \times 10^{7} ~ \rm yr$, which
is not much smaller than the time interval between the epoch at which
a suitably massive SMBH could first have formed and the end of the EoR
(Figure 6; \cite{Willott03}).  It is therefore reasonable to assume
that most SMBHs within the EoR will be observed accreting at some high
fraction of the Eddington-limited rate, although the possibility that
an important proportion of black hole growth occurs in short-lived
super-Eddington-accreting bursts \cite{Collin02} can never be easily
discounted. Since the SKA can penetrate the huge optical depths common
in the EoR, and since its will have unparalleled sensitivity to quasar
activity over wide fields of view (FOV), it will become the premier
instrument for understanding quasar (i.e.\ SMBH) feedback within the
EoR.

Some small fraction ($\sim 0.01$) of these accreting SMBH will
generate powerful radio jets (radio-loud quasars). Their radio
luminosities are immense so a significant fraction of those observable
in the EoR have probably already been detected in surveys like FIRST
and NVSS, and even shallow SKA surveys will trivially find them all.
However, \cite{Carilli03} have shown that high signal-to-noise ratio
spectroscopic observations are needed to measure reliable redshifts
for these objects (from the forest of HI 21cm absorption lines -- the
radio analogue of the Gunn-Peterson effect) and since they will be
rare, and very hard to find in other ways, they will perhaps be both
found and studied using an `all-hemisphere' SKA survey (figure 5).

\begin{figure} 
\includegraphics[width=3in]{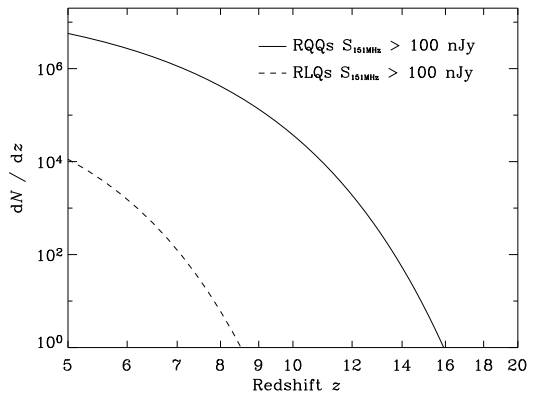}
\caption{The predicted differential number $\rm{d} N / \rm{d} z$ of
`radio-loud' quasars (dashed line) and `radio-quiet' QSOs (solid line)
in an all-hemisphere ($2 \pi$), 1-year SKA survey at $\approx 0.15$
GHz (assuming a $5 \sigma$ continuum detection limit of $\approx 100$
nJy, a $FOV = 1 ~ \rm deg^2$ at $\nu = 1.4$ GHz, scaling as
$\nu^{-2}$). The prediction follows from the methods outlined in
Jarvis \& Rawlings (this volume) and are consistent with the cosmic
evolution in AGN trigger rates inferred from Fig. 6.  All Radio-loud
quasars [defined here as all objects, including optically-obscured
`radio galaxies'], in the sub-population dominating the radio
luminosity function (RLF) at and above its characteristic break
\cite{Willott01} will be detected at $>$mJy levels and will
therefore have high signal-to-noise SKA spectra, allowing studies of
the IGM during the EoR using HI absorption \cite{Carilli03}.
The SKA sensitivity is also sufficient to detect all the radio-quiet
quasars in the EoR [defined here as all quasars, including
optically-obscured `Type-II' objects], at or above the characteristic
break in the X-ray luminosity function \cite{Ueda03}.
}
\end{figure}

\begin{figure} 
\includegraphics[width=3in]{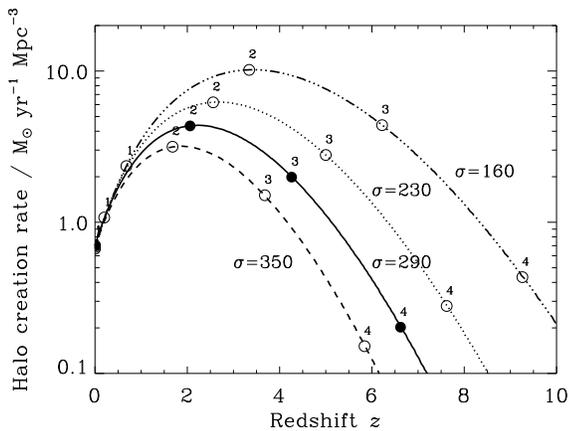}
\caption{The creation rate of dark-matter halos as a function of redshift $z$
for various halo velocity dispersions $\sigma$ (in units of $\rm km ~
s^{-1}$) from Rawlings \& Jarvis \cite{Rawlings04}. Adopting the
feedback model (for powerful jets) of Rawlings \& Jarvis, the halos
capable of hosting these `radio-loud' objects lie in the range $230
\leq \sigma \leq 350$, which correspond to $\nu=4$ (where $\nu$,
marked by circles, is the density threshold for collapse in units of
the r.m.s. density fluctuations) at $z \sim 6-8$.  Adopting the
feedback model (for weak jets) of Silk \& Rees (1998), the halos
capable of hosting these `radio-quiet' objects reach down to $\sigma =
160$, and hence can form from a $\nu=4$ fluctuation at $z \sim 9.5$.
}
\end{figure}

Carilli et al. \cite{Carilli03} have argued that studies of the IGM
within the EoR are made possible by HI 21cm absorption against distant
radio-loud sources.  Such objects are also important in themselves as
they probably contribute significantly to the heating of the IGM, its
magnetization and the distribution of heavy elements during the EoR.
Their influence on the ionized component of the cosmic web is likely
to be profound, e.g. \cite{Gopal01}, and their jets carry enough
kinetic power that they dominate the mechanical output of all
accreting black holes. A single radio-loud outburst can, for example,
shut-down star-formation and black-hole activity in $\sim 10-100$
surrounding protogalaxies in a protocluster \cite{Rawlings04}.

Radio-quiet quasars are the dominant source of photoionizing (if not
kinetic) power associated with accreting SMBH.  They always have some
radio emission, e.g.\ from star formation and weak jets. The SKA has
sufficient sensitivity to detect essentially all this population
within the EoR (Figure 6).  Although finding these objects represents
a challenging sifting problem, large FOV (c.f. present and future
X-ray and mid-IR satellites) and (via deep SKA exposures) redshift
measurement using associated HI 21cm (rather than metal) lines, may
prove crucial in finding the first (i.e.  highest-redshift) SMBHs in
the Universe.

\subsection{Compact radio sources}

In order to find the first SMBHs, besides its sensitivity and powerful
surveying capability, the SKA can make additional use of its resolving
power for identifying AGN. Almost every AGN, be it radio-loud or
radio-quiet, is associated with a high brightness temperature radio
core (this is discussed in more detail by Falcke, K\"ording, \& Nagar
in this volume).  The radio core is typically the inner part of a
moderately relativistic plasma jet. At flux density levels of several
mJy these radio cores are almost exclusively found to be associated
with the SMBH in a galaxy. They stand out when using a spatial filter
concentrating on the longest baselines of the telescope array. 

Radio cores have been used effectively to identify AGN in wide field
VLBI images of deep fields and for other AGN surveys. They are even
more pronounced for low-power sources. Surveying the Palomar sample of
nearby galaxies at a shallow level with the VLA, Nagar et
al. \cite{Nagar00,Nagar02} showed that $\sim40\%$ of nearby AGN host a
compact radio core that is easily detected and which reliably
identifies AGN activity. At lower resolution, including extended
emission, the detection rate even approaches $85\%$ \cite{Ho01}.

Compact radio core emission seems to be a function of accretion power
and is related to the X-ray emission and the black hole mass. Black
holes -- stellar and supermassive -- form a loosely defined
fundamental plane in the parameter space opened up by radio flux,
X-ray flux, and black hole mass
\cite{Falcke04,Merloni03}. This relation implies
that supermassive black holes are most effectively detected in the
radio while X-rays are more effective for stellar mass black holes.

\begin{figure}[ht] \resizebox{\columnwidth}{!}{
\includegraphics{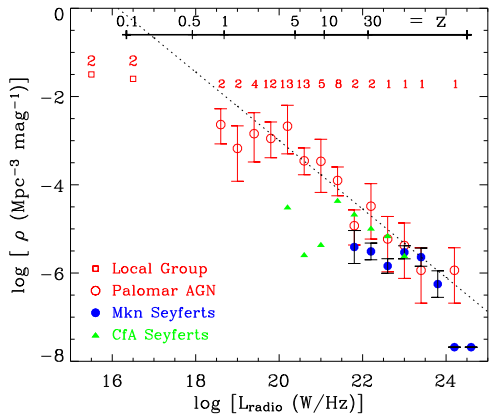} } 
\caption{The 15~GHz nuclear radio
luminosity function (RLF) of the 150~mas-scale radio nuclei in the
Palomar sample (red open circles, with the number of galaxies in each
bin listed above the symbol) as compared to the RLFs for Markarian
Seyferts and CfA Seyferts.  The dashed line is a power-law ($-0.78$)
fit to the Palomar RLF.  Also shown is the estimated nuclear RLF of
galaxies in the local group.  The upper scale illustrates the
detection limits of the SKA (using an r.m.s noise of 0.3$\mu$Jy in
1~hr of imaging, and a 3$\sigma$ detection threshold) of similar AGNs
at redshifts of $z=0.1$--30 (Figure from \cite{Nagar04}, see also
Falcke, K\"ording, Nagar, this volume).}  \label{figrlf} 
\end{figure}

Figure \ref{figrlf} shows the local radio core luminosity function of
AGN (most of which are traditionally classified as radio-quiet) from
Seyferts galaxies, via Low-Luminosity AGN down to our own Milky
Way. The scale on the top shows the typical redshift range detectable
for such AGN with the SKA in one hour.  It shows that the SKA will
allow the study of a very wide range of AGN activity with
low-luminosity AGN being easily visibly out to z=1 and more luminous
ones even further. Besides studying AGN evolution, the SKA will thus
also penetrate deep into the EoR. Even without beaming, radio cores
are expected to be visible at $z>10$ from $10^7 M_\odot$ black holes
shining at their Eddington limit or from $10^9 M_\odot$ black holes
shining at 0.1\% of the Eddington limit. This allows one to detect the
very first SMBHs already in the process of their formation -- if this
happens through baryonic accretion.

The compact radio emission may, however, be further amplified if one
does not only consider the inner, fairly dissipationless part of the
jet. One could argue that the terminus of jets, i.e., their hot spots,
from the first black holes will be confined to the host galaxy due to
large ambient densities and pressures in the violent formation process
of the early galaxies. Since the hot spots are working surfaces of the
jet, their emission can dominate the overall radio emission easily and
the compactness of the emission makes them appear as high-brightness
temperatures sources. 

\begin{figure}
\centerline{\psfig{figure=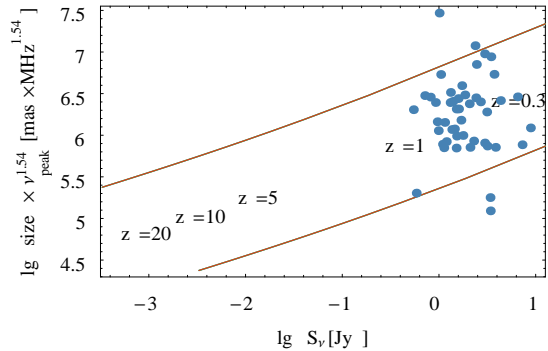,width=0.49\textwidth}}
\vspace{-1cm} \caption{\label{sizeandnu} Size, frequency, and flux
density roughly form a fundamental plane for GPS radio galaxies. Here
we collapse this plane by plotting a combination of frequency and size
($\nu_{\rm peak}^{1.54}\times r$) vs. the flux density for a range of
empirical GPS models and actual GPS sources. The two lines delimit the
scatter in the empirical relation. The labels indicate the typical
redshift of sources in that regime. Standard GPS sources are found in
the redshift range around z=1. The bottom left corner is not occupied
and is the discovery space for young high-redshift black holes with
the SKA (Figure from Falcke, K\"ording, Nagar, this volume).}
\end{figure}

The radio cores will then have a very characteristic, peaked broad
band spectrum (called Gigahertz-Peaked Spectrum/GPS
sources\cite{O'Dea1998}) caused by synchrotron absorption with a
double-lobed spatial structure on a scale of some tens of
milli-arcsecond. This characteristic peak allows for designing very
effective large survey capabilities as the physical size and the peak
frequency are related: the smaller the source the higher the
frequency. Fortuitously frequency and angular size show a very
different behaviour as one goes to large redshifts: frequency will
decrease while angular sizes become larger again for a standard ruler
at $z>1$. Hence in the parameter space of angular size, peak
frequency, and flux density, ultra-high redshift GPS sources should
occupy a unique region (lower left corner in
Fig.~\ref{sizeandnu}). They will have peak frequencies in the 0.1-0.3
GHz regime, sizes of some 10-100 milli-arcsecond, and flux densities
in the mJy regime. 

An all-sky broad-band survey with high-resolution follow-up
observations should therefore very effectively produce candidates for
the first SMBHs in the universe. Some of them may even be bright
enough to do HI absorption spectroscopy in deep integrations to
confirm their high-z nature. The crucial and qualitatively new features
of the SKA to make this succeed are its huge instantaneous band-width
to measure spectral curvature and its multi-beaming capability to
facilitate large surveys.

In summary, the SKA seems to be an ideal instrument to study AGN
evolution at all luminosity levels and all redshifts largely
unaffected by dust obscuration. It even provides a number of angles on
how to identify and study in detail the very first generation of
SMBHs. As such the SKA will be an invaluable tool for cosmology and
black hole physics alike.

\end{document}